\documentclass[aps,prl,superscriptaddress,twocolumn,showpacs,amsmath,amssymb,reprint,longbibliography,floatfix]{revtex4-2}

\usepackage{graphicx}
\usepackage[percent]{overpic}
\usepackage{placeins}
\usepackage{dcolumn}
\usepackage{bm}
\usepackage[T1]{fontenc}
\usepackage[utf8]{inputenc}
\usepackage{xcolor}
\usepackage[colorlinks=true,allcolors=black]{hyperref}

\usepackage{amsfonts}
\usepackage{mathptmx}
\begin{document}

\title{{One-Dimensional Type-IV Collinear Magnetism}}

\author{Xin Chen}
\email{xin.chen@iat.cn}
\affiliation{Thermal Science Research Center, Shandong Institute of Advanced Technology, Jinan 250100, Shandong Province, People's Republic of China}


\author{Duo Wang}
\affiliation{Faculty of Applied Sciences, Macao Polytechnic University, Macao SAR, 999078, People's Republic of China}

\author{Peng Zhang}
\affiliation{Thermal Science Research Center, Shandong Institute of Advanced Technology, Jinan 250100, Shandong Province, People's Republic of China}



\author{Zeying Zhang}
\email{zzy@buct.edu.cn}
\affiliation{College of Mathematics and Physics, Beijing University of Chemical Technology, Beijing 100029, China}


\date{\today}

\begin{abstract}
{Two-dimensional type-IV collinear magnets combine nonrelativistic spin degeneracy with spin-orbit-coupling-enabled magnetic responses. Here we establish a one-dimensional counterpart through altermagnetic confinement. A Lieb-based construction connects three-dimensional altermagnets to two-dimensional type-IV magnets and extends the same principle to one-dimensional ribbons. In a symmetric $[110]$ ribbon, a sublattice-exchange operation preserves the longitudinal momentum and enforces nonrelativistic spin degeneracy. {Spin-orbit coupling (SOC) ties spin to the lattice, so the N\'eel orientation determines whether the remaining symmetries protect the doublets or allow them to split.} First-principles-derived Wannier calculations for Ta$_2$TeSeO ribbons reveal spin-polarized propagating channels with an ideal right-moving-mode polarization of approximately $35\%$ at finite conductance for the parent easy-axis orientation $\mathbf N\parallel[100]$. Direct DFT calculations of Ta$_2$Te$_2$O ribbons resolve the contrast between spin-split $[100]$ and spin-degenerate $[001]$ SOC spectra. Assembling the same isolated Ta$_2$Te$_2$O ribbons into a transversely periodic array restores spin splitting without SOC. This conversion parallels the recovery of three-dimensional altermagnetism through suitable periodic stacking of two-dimensional type-IV magnets. One-dimensional type-IV magnetism thus provides a route to spin splitting through the interplay of antiferromagnetic order and SOC, distinct from conventional Rashba splitting driven by structural inversion asymmetry.}
\end{abstract}

\maketitle

The electronic classification of collinear magnets depends not only on their magnetic order, but also on how the symmetries relating opposite-spin sublattices act in momentum space{~\cite{PhysRevX.12.040501,Jungwirth2025,Jungwirth2026,Sato2026Octupole}}.
Altermagnetism makes this distinction explicit: a compensated collinear structure can exhibit nonrelativistic spin-split bands{~\cite{Hayami2019Momentum,PhysRevB.102.014422,Yuan2021SpinSplittingPrototypes,PhysRevB.102.144441,PhysRevLett.132.236701,Manc2021,10.1021/acs.nanolett.6c01573,zhiheng2025spinaxisdynamiclocking}} when crystalline rotations connect its opposite-spin sublattices and exchange inequivalent momenta{~\cite{smejkal2020crystal,PhysRevX.12.031042,Smolyanyuk2024Amcheck}}.
Recent studies of two-dimensional (2D) magnets have revealed a further possibility. Type-IV collinear magnets retain nonrelativistic spin degeneracy within an isolated layer. Suitable periodic stacking can recover three-dimensional altermagnetism with nonrelativistic spin splitting. Within the isolated layer, spin-orbit coupling (SOC) can also lift the degeneracy, depending on the N\'eel-vector orientation~\cite{Bai2025TypeIV,jp95-17sz}. {This distinction raises a natural question about dimensionality. Can the same magnetic principle be realized in one dimension, where only the longitudinal Bloch momentum remains?}

{Dimensional confinement offers a physical route to this one-dimensional counterpart. In an altermagnet, crystalline operations connect opposite-spin sublattices at different momenta~\cite{PhysRevX.12.031042}. A symmetry-preserving cut can retain the sublattice relation while making the same operation leave the remaining momentum unchanged. Such a construction connects a three-dimensional (3D) altermagnet to the spin-degenerate spectrum characteristic of a 2D type-IV magnet~\cite{Bai2025TypeIV,jp95-17sz}. Applying this principle to a 2D altermagnet suggests a 1D realization. The essential requirement is that the finite structure, including its boundaries, preserve the relevant sublattice-exchange operation.}

{A diagonal ribbon of a $d$-wave Lieb-type altermagnet provides a concrete setting~\cite{Manc2021,10.1063/5.0147450,v38b-5by1,Xu2025Alterpiezoresponse}. The parent bands are spin split along the principal axes but degenerate along $[110]$~\cite{PhysRevB.111.134429,Khan2025}. A symmetry-compatible ribbon along this diagonal extends the degeneracy to the entire nonrelativistic subband spectrum. This geometry provides a symmetry-protected, spin-degenerate nonrelativistic starting point, distinguishing the spin splitting that emerges with SOC from the nonrelativistic splitting of the parent altermagnet. This differs from a $\mathcal{PT}$-symmetric wire, in which combined inversion and time reversal continue to protect degeneracy with SOC~\cite{PhysRevB.102.014422,Yuan2021SpinSplittingPrototypes}. The distinction makes the diagonal ribbon a direct setting for establishing the one-dimensional counterpart of type-IV magnetism.}

\begin{figure*}[htbp]
\centering\includegraphics[width=17.78cm]{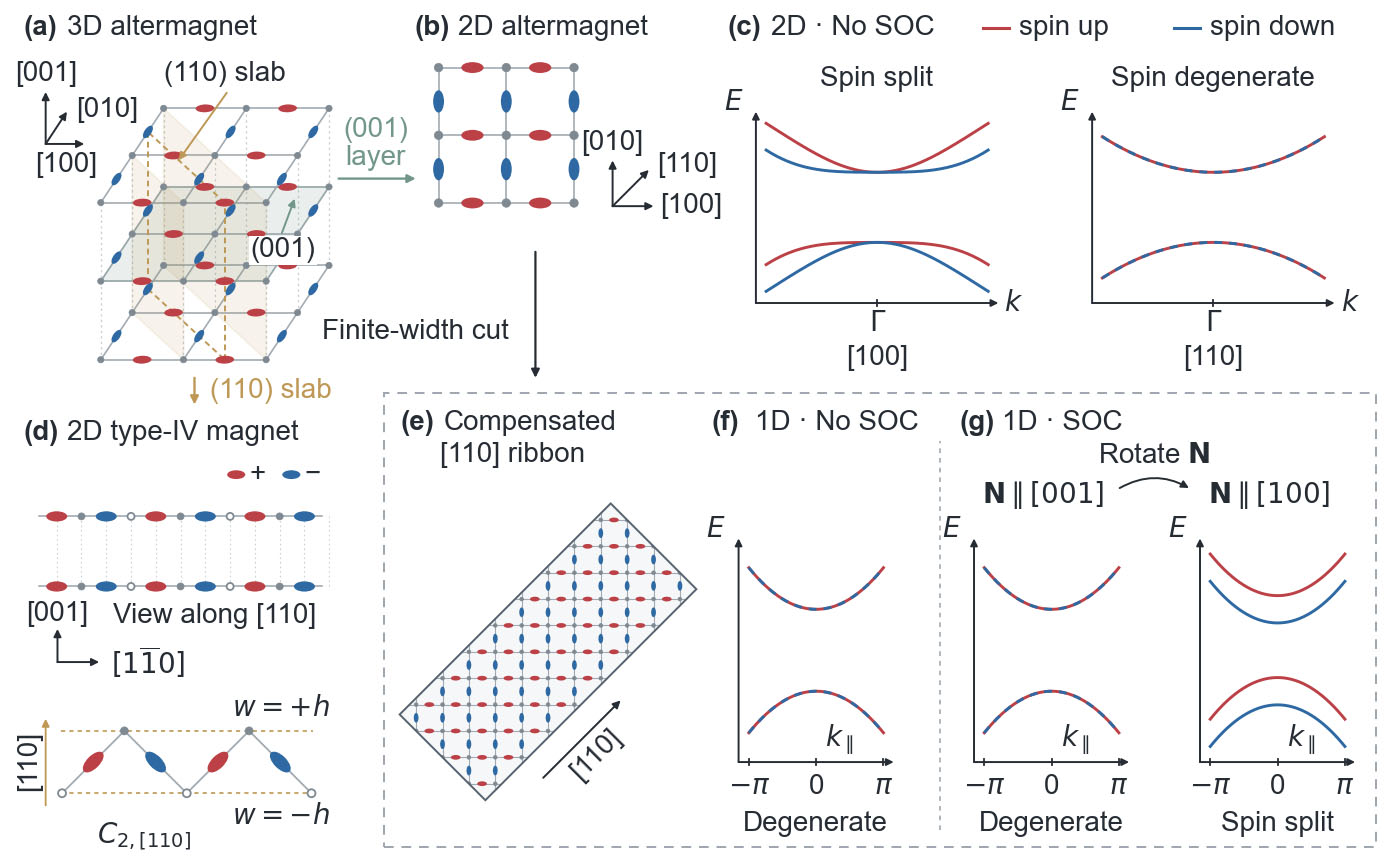}
\caption{{Dimensional connection between altermagnetism and type-IV magnetism in an ideal Lieb-based construction.} (a) AA-stacked magnetic Lieb parent, with the selected $(001)$ layer and $(110)$ slab highlighted. (b) Two-dimensional Lieb altermagnet. (c) Nonrelativistic dispersions showing spin splitting along $[100]$ and degeneracy along $[110]$. (d) $(110)$ slab realizing the type-IV limit through the spin-exchanging $C_{2,[110]}$ symmetry; solid and open gray sites mark the two nonmagnetic sublayers. {(e) $[110]$ ribbon with opposite-spin sublattices and symmetry-related edges.} (f) Non-SOC ribbon bands remain spin degenerate. {(g) In the horizontal-mirror-symmetric ribbon, SOC preserves the $[001]$ doublets, while rotation to $[100]$ allows spin splitting.} Red and blue denote opposite spin sectors or projected spin character; gray sites are nonmagnetic. All dispersions are schematic.}
\label{fig:mechanism}
\end{figure*}

{In this Letter, we establish one-dimensional type-IV magnetism through altermagnetic confinement. A Lieb-based construction relates the 3D-to-2D and 2D-to-1D reductions in Fig.~\ref{fig:mechanism}. We then demonstrate the 1D realization using Wannier Hamiltonians and spin operators for symmetric $[110]$ Ta$_2$TeSeO ribbons, derived from the altermagnetic monolayer~\cite{ChenTa2TeSeO2026}. Their nonrelativistic bands are spin degenerate, whereas SOC produces N\'eel-orientation-dependent spectra and spin-polarized propagating channels. For $\mathbf N\parallel[100]$, a parent easy-axis orientation, the ideal right-moving-mode polarization reaches approximately $35\%$ at finite conductance. Direct DFT calculations of non-Janus Ta$_2$Te$_2$O ribbons independently demonstrate the contrast between spin-split $[100]$ and spin-degenerate $[001]$ SOC states. Transverse assembly of the same Ta$_2$Te$_2$O ribbons restores spin splitting without SOC, paralleling the recovery of 3D altermagnetism by suitable periodic stacking of 2D type-IV magnets. Together, these results extend the type-IV mechanism to one dimension and identify its longitudinal spin-transport manifestation.}

The dimensional construction in Fig.~\ref{fig:mechanism} begins with an ideal AA-stacked three-dimensional parent made of magnetic Lieb layers [Fig.~\ref{fig:mechanism}(a)]. Selecting a single $(001)$ layer gives a two-dimensional $d$-wave Lieb altermagnet [Fig.~\ref{fig:mechanism}(b)]. In this layer, the sublattice-exchange operation swaps the two in-plane momentum components,
\begin{equation*}
E_\uparrow(k_x,k_y)=E_\downarrow(k_y,k_x),
\end{equation*}
so the leading nonrelativistic splitting has the form $\Delta E\propto k_x^2-k_y^2$. The spectrum is spin split along the principal directions $[100]$ and $[010]$, but remains degenerate along the diagonal $[110]$ [Fig.~\ref{fig:mechanism}(c)]~\cite{PhysRevX.12.031042,bf1n-sxdl}.

{A finite $(110)$ slab of the same parent gives the 2D type-IV construction in Fig.~\ref{fig:mechanism}(d). The spin-exchanging $C_{2,[110]}$ operation relates the two spin sectors, whose spectra coincide in this nonrelativistic slab construction. The corresponding 1D construction is a finite-width $[110]$ strip of the Lieb layer with symmetry-related edges [Fig.~\ref{fig:mechanism}(e)].} {The vertical mirror $m_v=m_{(1\bar10)}$, whose plane contains the ribbon axis $[110]$ and $[001]$, exchanges the opposite-spin sublattices and the two edges while preserving $k_\parallel$. Taking $g=m_v$ and denoting the collinear exchange field by $b(\mathbf r)$ gives}
\begin{equation}
b(g\mathbf r)=-b(\mathbf r),\quad gk_\parallel=k_\parallel
\ \Longrightarrow\ E_\uparrow(k_\parallel)=E_\downarrow(k_\parallel).
\label{eq:1d_degeneracy}
\end{equation}
{Without SOC, spin rotations are independent of the lattice. Combining $g$ with a $\pi$ spin rotation perpendicular to $\mathbf N$ maps each spin-up eigenstate onto an orthogonal spin-down partner at the same $k_\parallel$ and energy~\cite{ChenX2025,PhysRevX.14.031037}. This relation holds for every transverse subband and every N\'eel orientation, enforcing spin degeneracy throughout the ribbon Brillouin zone [Fig.~\ref{fig:mechanism}(f)]. The finite ribbon therefore inherits a symmetry-protected degeneracy of its complete spectrum.}

{SOC ties spin rotations to the lattice, so the N\'eel orientation determines which operations remain magnetic symmetries~\cite{PhysRevX.14.031037,jp95-17sz}. In the horizontal-mirror-symmetric ribbon of Fig.~\ref{fig:mechanism}(g), $\mathbf N\parallel[001]$ preserves both $m_v$ and $m_z=m_{(001)}$. Both mirrors leave $k_\parallel$ unchanged. Their spin-$1/2$ representations anticommute, so applying one mirror to an eigenstate of the other produces an orthogonal partner with the same energy and the opposite mirror eigenvalue. This protects doublets throughout the one-dimensional Brillouin zone. Rotating $\mathbf N$ to $[100]$ removes this pair of unitary mirrors and permits SOC splitting. The orientation dependence thus follows from the magnetic symmetries retained after spin and lattice rotations become coupled. { This mechanism differs from conventional Rashba splitting, which requires structural inversion asymmetry but no magnetic order~\cite{Manchon2015Rashba}. Here, SOC lifts a degeneracy protected by the nonrelativistic antiferromagnetic symmetry, with the surviving magnetic symmetries determining its dependence on the N\'eel orientation.} The representation analysis and a $\mathcal{PT}$-protected tight-binding reference are given in Supplemental Material (SM)~\cite{SupplementalMaterial}. These symmetry requirements concern the final one-dimensional structure and do not require a higher-dimensional parent.}

\begin{figure*}[tbp]
\centering
\includegraphics[width=17.78cm]{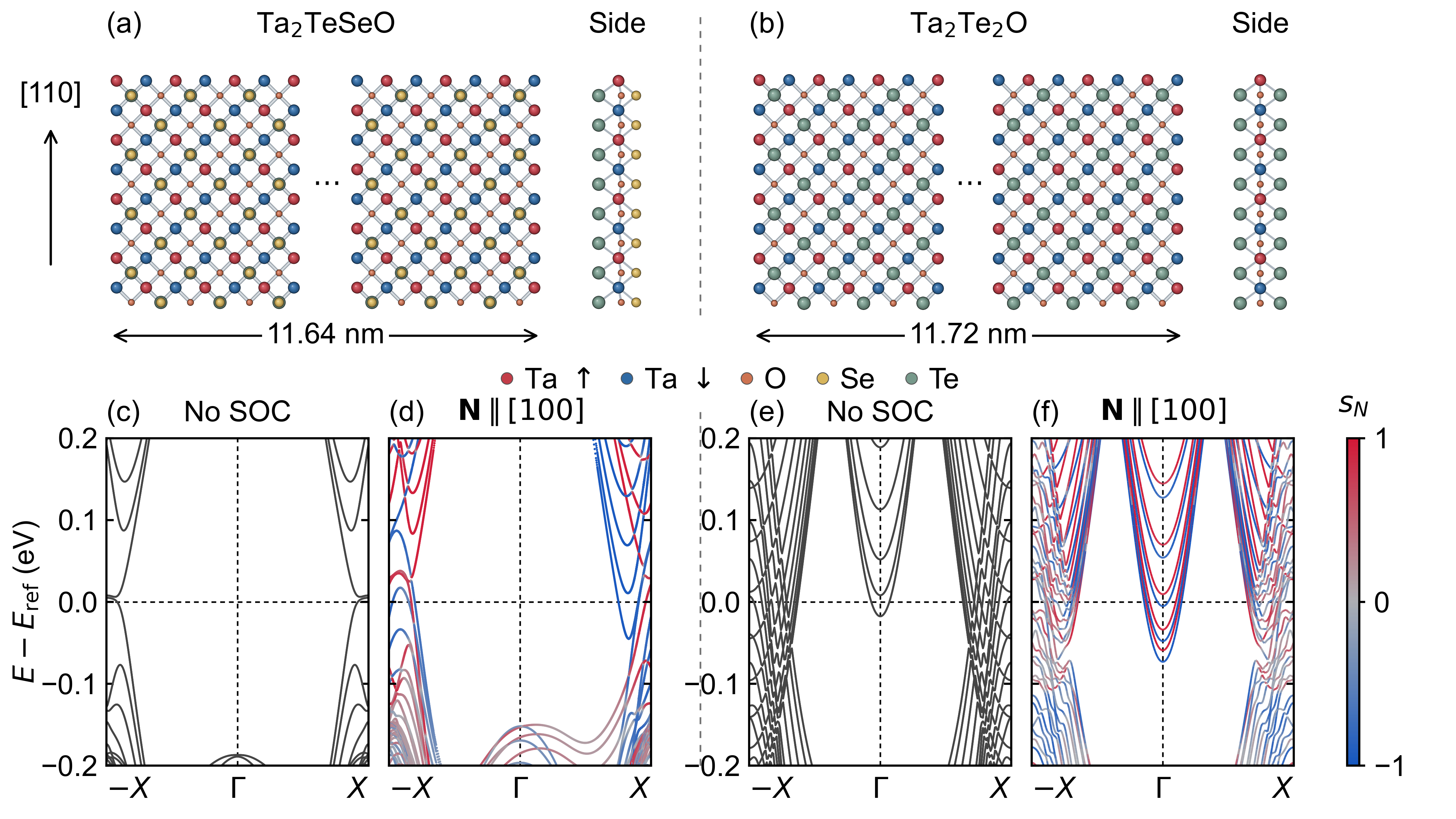}
\caption{{SOC lifts the nonrelativistic spin degeneracy in both Janus Ta$_2$TeSeO and non-Janus Ta$_2$Te$_2$O $[110]$ ribbons. (a),(b) Top and side views, with the middle of each ribbon omitted in the top view. (c),(d) Ta$_2$TeSeO bands without SOC and with SOC for $\mathbf N\parallel[100]$. (e),(f) Corresponding Ta$_2$Te$_2$O bands. Dark gray in (c),(e) shows coincident opposite-spin spectra. SOC colors give $s_N$ along the N\'eel axis, with light gray indicating weak projection. Dashed lines mark $\Gamma$ and $E_{\rm ref}$, the Fermi energy of each parent SCF calculation.}}
\label{fig:ta2teseo_bands}
\end{figure*}

{We test the material realization in Janus Ta$_2$TeSeO and non-Janus Ta$_2$Te$_2$O ribbons (Fig.~\ref{fig:ta2teseo_bands}). The Ta$_2$TeSeO monolayer contains a square Ta network decorated by Te, Se, and O, with nonmagnetic space group $P4mm$ (No.~99) and point group $C_{4v}$~\cite{ChenTa2TeSeO2026}. The side views in Figs.~\ref{fig:ta2teseo_bands}(a,b) show the asymmetric Te/Se environment in Ta$_2$TeSeO and the equivalent Te layers in Ta$_2$Te$_2$O. The Janus polarity removes the horizontal mirror. The Ta$_2$TeSeO parent bands are spin split along the principal axes and degenerate along the diagonals without SOC, and its magnetic easy axes lie along $[100]$ and $[010]$~\cite{ChenTa2TeSeO2026}.}

{Both ribbons are periodic along $[110]$ and finite along $[1\bar10]$, with the same cut and number of transverse repeats [Figs.~\ref{fig:ta2teseo_bands}(a,b)]. The mirror-symmetric Ta-terminated Ta$_2$TeSeO ribbon spans $11.64~\mathrm{nm}$ and contains 197 atoms per axial period, with 40 Ta sites in each magnetic sublattice. The non-Janus Ta$_2$Te$_2$O ribbon spans $11.72~\mathrm{nm}$. Their symmetry-related edges preserve the sublattice-exchange mapping in Eq.~\eqref{eq:1d_degeneracy}, and both have spin-degenerate non-SOC bands [Figs.~\ref{fig:ta2teseo_bands}(c,e)].}

{The first-principles input is obtained using the projector augmented-wave method~\cite{PhysRevB.50.17953} implemented in VASP~\cite{PhysRevB.54.11169}. We construct the ribbon Hamiltonians by cutting the monolayer Wannier Hamiltonians into finite strips~\cite{MarzariVanderbilt1997,SouzaMarzariVanderbilt2001,Pizzi2020Wannier90}. With SOC, the Hamiltonian and spin Pauli vector $\boldsymbol\sigma$ are transformed into the same Wannier gauge. The band colors give the spin expectation value along the N\'eel vector,
\begin{equation}
s_N(n,k_\parallel)=\langle\psi_{nk_\parallel}|\boldsymbol\sigma\cdot\hat{\mathbf N}|\psi_{nk_\parallel}\rangle,
\label{eq:physical_spin_projection}
\end{equation}
where $|\psi_{nk_\parallel}\rangle$ is a ribbon eigenstate and $\hat{\mathbf N}$ is the unit N\'eel vector. Red and blue denote positive and negative $s_N$, respectively. SOC mixes spin, so the colors represent spin expectation values rather than conserved spin-channel occupations. For Ta$_2$TeSeO, independent SOC Wannier Hamiltonians are used for $[100]$, $[001]$, and $[110]$, with the $[1\bar10]$ case obtained through the $C_{4z}$-related construction. Benchmarks against monolayer DFT confirm that the Wannier model reproduces the low-energy dispersions and spin character. The benchmarks, full orientation comparison, and computational details are given in the SM~\cite{SupplementalMaterial}.}

{For $\mathbf N\parallel[100]$, both ribbons develop clearly separated branches with opposite signs of $s_N$ at the same momentum [Figs.~\ref{fig:ta2teseo_bands}(d,f)], lifting their non-SOC doublets [Figs.~\ref{fig:ta2teseo_bands}(c,e)]. The non-Janus Ta$_2$Te$_2$O result demonstrates that this SOC-induced splitting does not require Janus asymmetry. In Ta$_2$TeSeO, $[100]$ is a parent easy-axis direction, making the spin-split state an equilibrium candidate for sufficiently wide ribbons that retain the parent magnetic anisotropy. In this regime, the splitting requires no externally imposed N\'eel-vector rotation. The self-consistent Ta$_2$TeSeO SOC reference also carries a finite net spin moment alongside this band reconstruction.}

{Table~\ref{tab:janus_rod_symmetry} lists the magnetic rod groups and double-valued small corepresentations of the Janus ribbon at $\Gamma$, a generic point $\Lambda$, and $X$. The dimension is given in parentheses. A two-dimensional corepresentation enforces a doublet, whereas two distinct one-dimensional corepresentations need not have the same energy. These groups contain the spatial operations, combined with time reversal when required, that preserve the one-dimensionally periodic magnetic structure. For $\mathbf N\parallel[100]$, the remaining antiunitary operation $C_{2z}\mathcal T$, where $\mathcal T$ is time reversal, squares to $+1$ and does not enforce a doublet, consistent with Fig.~\ref{fig:ta2teseo_bands}(d). For $\mathbf N\parallel[001]$, the ribbon retains two vertical mirrors, $m_v$ and $m_u=m_{(110)}$. The second is normal to the ribbon and reverses $k_\parallel$, so both act at the same momentum only at $\Gamma$ and $X$, where their anticommuting spinor representations protect doublets. Splitting is allowed at general momenta, and the two diagonal in-plane orientations also allow nondegenerate bands.}

\begin{table*}[t]
\caption{Magnetic rod groups and double-valued small corepresentations of the $[110]$ Ta$_2$TeSeO ribbon with SOC. Parentheses give representation dimensions. Identical subscripts in different groups do not denote the same representation.}
\label{tab:janus_rod_symmetry}
\begin{ruledtabular}\begin{tabular}{ccccc}
$\mathbf N$ & Magnetic rod group & $\Gamma$ & Generic point $\Lambda$ & $X$\\
$[100]$ & $p2'11$ & $\Gamma_2(1)$ & $\Lambda_2(1)$ & $X_2(1)$\\
$[001]$ & $p2mm$ & $\Gamma_5(2)$ & $\Lambda_3(1),\Lambda_4(1)$ & $X_5(2)$\\
$[110]$ & $p2'mm'$ & $\Gamma_3(1),\Gamma_4(1)$ & $\Lambda_3(1),\Lambda_4(1)$ & $X_3(1),X_4(1)$\\
$[1\bar10]$ & $p2'm'm$ & $\Gamma_3(1),\Gamma_4(1)$ & $\Lambda_2(1)$ & $X_3(1),X_4(1)$\\
\end{tabular}\end{ruledtabular}
\end{table*}

{The absence of the horizontal mirror $m_z$ in Ta$_2$TeSeO distinguishes its $[001]$ spectrum from that of the ideal ribbon in Fig.~\ref{fig:mechanism}(g). The non-Janus Ta$_2$Te$_2$O ribbon restores this mirror, recovering full-path doublets through the $m_v$--$m_z$ pair. In the Janus $[001]$ domain, $m_v$ reverses the out-of-plane spin but preserves its transverse component. A nondegenerate mirror eigenstate can therefore carry transverse spin with zero $s_N$, so weak N\'eel-axis projection does not imply energy degeneracy. The Ta$_2$TeSeO $[001]$ and diagonal-domain spin textures are given in the SM, together with the complete spin-rod-group and magnetic-rod-group analysis, operation matrices, and degeneracy proofs underlying Table~\ref{tab:janus_rod_symmetry}~\cite{SupplementalMaterial}.}

\begin{figure*}[tbp]
\centering
\includegraphics[width=17.78cm]{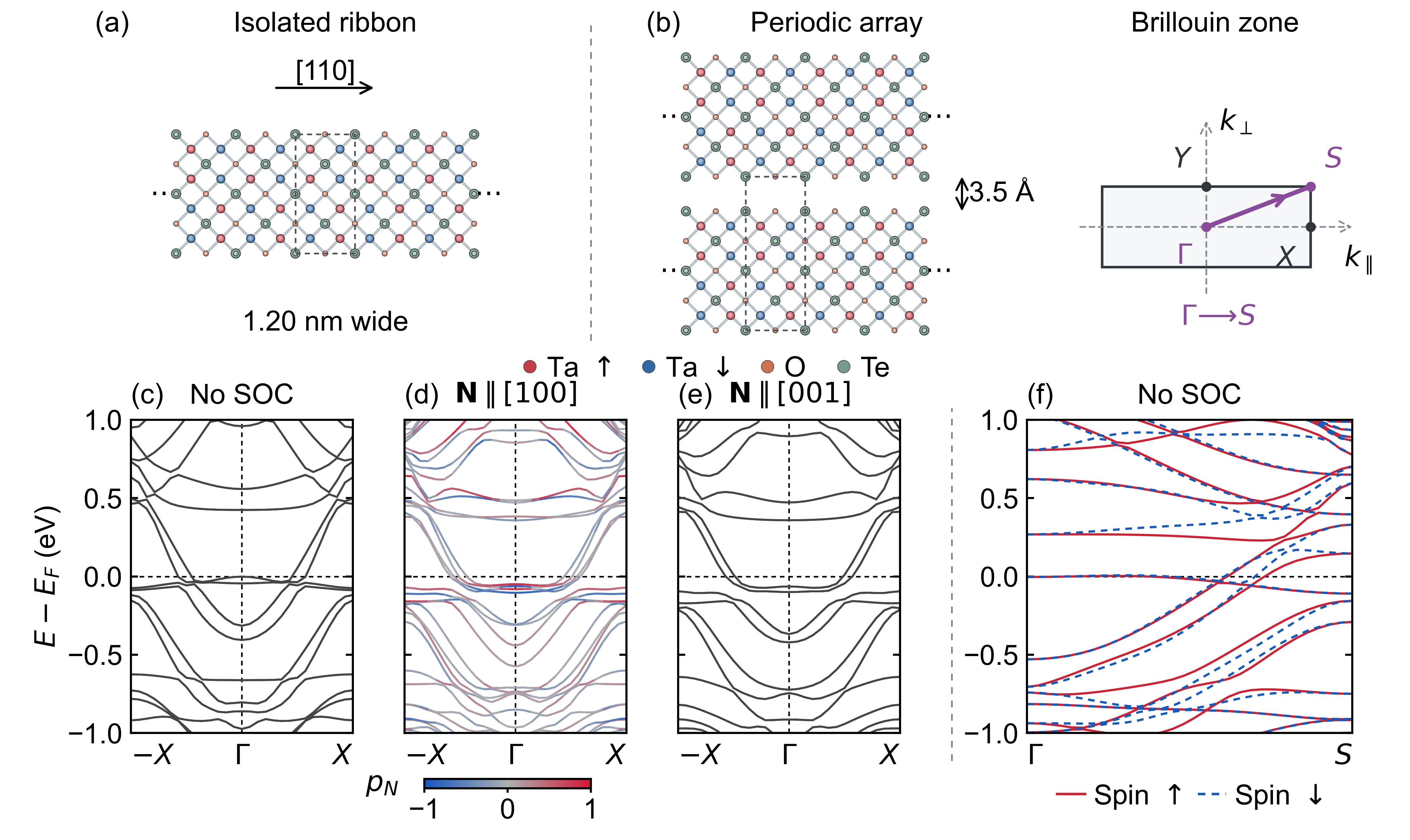}
\caption{{N\'eel-orientation-dependent degeneracy and recovery of nonrelativistic spin splitting in Ta$_2$Te$_2$O. (a),(b) Top views of the isolated ribbon and its transverse periodic array with a 3.5~\AA\ edge separation. Concentric Te symbols represent sites at opposite heights. The array Brillouin zone highlights the $\Gamma$--$S$ path, with $k_\parallel$ along the ribbon and $k_\perp$ transverse to it. (c--e) Direct DFT bands of the isolated ribbon along $-X$--$\Gamma$--$X$ without SOC and with SOC for $\mathbf N\parallel[100]$ and $[001]$. Dark-gray curves in (c),(e) show degenerate pairs. Color in (d) gives the normalized PAW spin projection $p_N$ along $\mathbf N$. (f) Non-SOC array bands along $\Gamma$--$S$, with opposite spins in red solid and blue dashed lines. Energies use each calculation's SCF Fermi level. Enlarged comparisons are provided in the SM.}}
\label{fig:ta2te2o_dft_array}
\end{figure*}

{N\'eel-vector rotation tests the predicted doublet protection in the horizontal-mirror-symmetric Ta$_2$Te$_2$O ribbon. Direct DFT calculations use a 1.20-nm-wide, Te/O-terminated $[110]$ ribbon with the same fixed geometry for all three magnetic states [Fig.~\ref{fig:ta2te2o_dft_array}(a)]. Its non-SOC bands are spin degenerate [Fig.~\ref{fig:ta2te2o_dft_array}(c)]. With SOC, the $[100]$ state exhibits resolved splitting [Fig.~\ref{fig:ta2te2o_dft_array}(d)], including a 23.2~meV separation between two branches near $E_F$ at $\Gamma$. Rotating $\mathbf N$ to $[001]$ restores pairwise energy degeneracy throughout the displayed $E_F\pm1$~eV window along the full ribbon path [Fig.~\ref{fig:ta2te2o_dft_array}(e)], realizing the protection by $m_v$ and $m_z$. This orientation contrast is obtained with self-consistent edge-charge redistribution and independently of the Wannier construction. { The resolved splitting at $\Gamma$ also distinguishes this magnetic response from conventional time-reversal-symmetric Rashba splitting, which preserves Kramers degeneracy there~\cite{Manchon2015Rashba}. Its appearance and removal upon rotating $\mathbf N$ at fixed geometry demonstrate the role of magnetic symmetry rather than Janus polarity.} The unconstrained $[100]$ solution carries a net spin moment of $0.223\,\mu_{\rm B}$ per period. Enlarged bands and the $[001]$ comparison for the wide Wannier ribbon are given in the SM~\cite{SupplementalMaterial}.}

{Transverse coupling provides a second route to lifting the ribbon degeneracy, now without SOC. We arrange the same ribbons into a 2D periodic array with a 3.5~\AA\ edge-to-edge separation, keeping the intraribbon geometry fixed [Fig.~\ref{fig:ta2te2o_dft_array}(b)]. The highlighted $\Gamma$--$S$ path traverses both longitudinal and transverse momenta in the array Brillouin zone. Without SOC, the isolated ribbon is spin degenerate, whereas the array recovers spin-split bands at intermediate momenta along this path [Fig.~\ref{fig:ta2te2o_dft_array}(f)], including a 131~meV opposite-spin energy separation in the occupied bands. This ribbon-to-array conversion is the one-dimensional counterpart of recovering 3D altermagnetism by suitable periodic stacking of 2D type-IV magnets~\cite{Bai2025TypeIV,jp95-17sz}. In both constructions, the additional Bloch momentum allows the spin-exchanging operation to connect distinct momenta, permitting nonrelativistic spin splitting when the layers or ribbons are coupled. The calculation details and enlarged band comparisons are given in the SM~\cite{SupplementalMaterial}.}

\begin{figure}[htbp]\centering
\includegraphics[width=8.3cm]{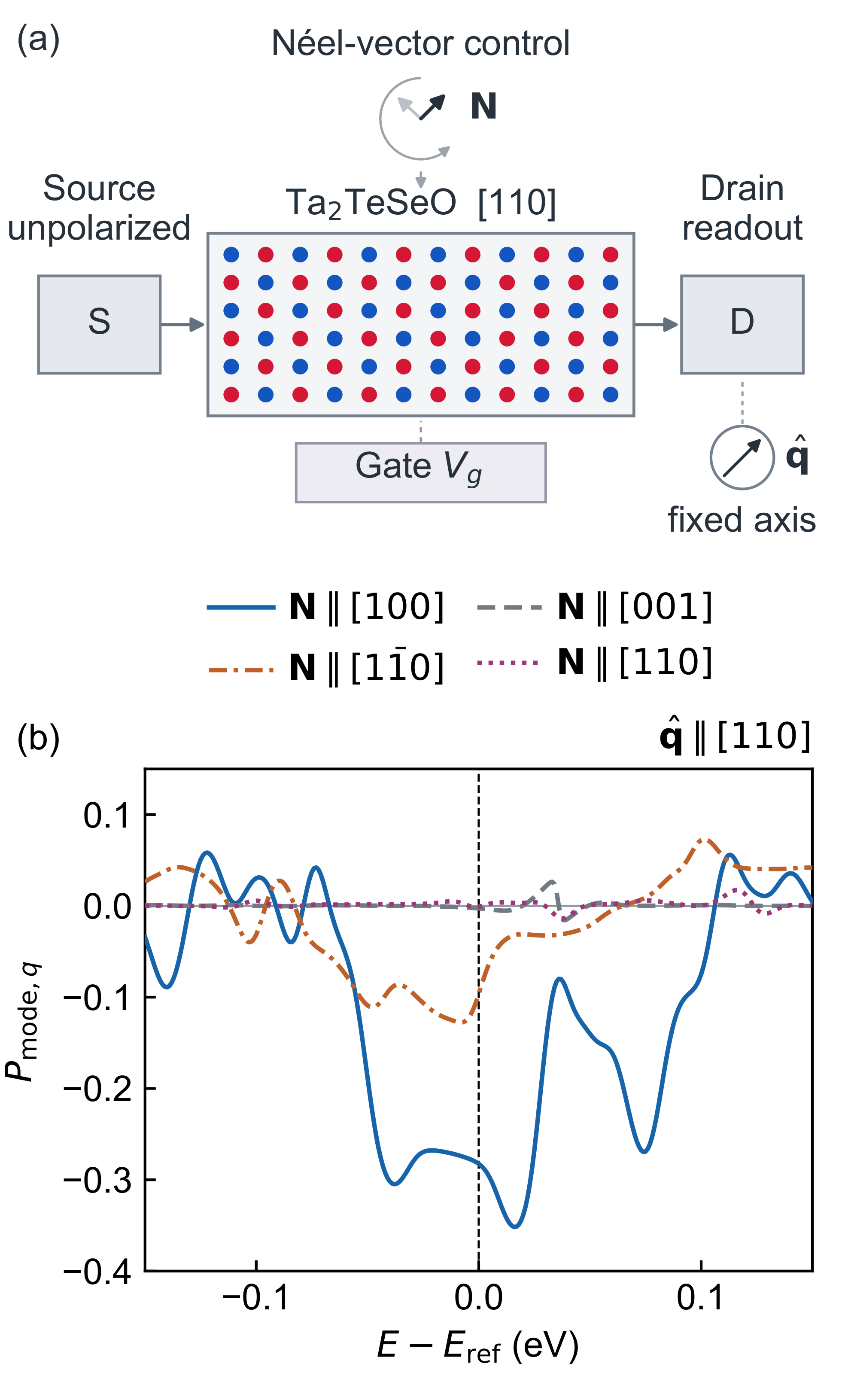}
\caption{{Net polarization of right-moving modes in the mirror-symmetric Ta-terminated $[110]$ Ta$_2$TeSeO ribbon. (a) Gated two-terminal readout with fixed spin-analysis axis $\hat{\mathbf q}\parallel[110]$. (b) Unit-transmission mode polarization $P_{{\rm mode},q}(E)$ for $\mathbf N\parallel[100]$, $[001]$, $[1\bar10]$, and $[110]$ in blue solid, gray dashed, orange dash-dotted, and purple dotted lines. Ideal conductance and numerical details are given in the SM.}}
\label{fig:transport_readout}
\end{figure}

{Oppositely polarized branches can cancel when all propagating channels contribute to transport. We therefore evaluate whether a net spin polarization survives the sum over all right-moving modes~\cite{PhysRevX.12.011028,PhysRevLett.128.197202}. Figure~\ref{fig:transport_readout}(a) sketches a gated two-terminal readout~\cite{10.1093/nsr/nww026}. The calculation uses a unit-transmission mode average with a fixed spin-analysis axis $\hat{\mathbf q}\parallel[110]$, comparing the same spin component for every N\'eel orientation.}

For a unit-transmission benchmark, let $\mathcal R(E)$ denote the set of right-moving Bloch modes at energy $E$, namely states with positive group velocity along the ribbon, and let $N_+(E)=|\mathcal R(E)|$. For each mode $\nu$, $s_{\nu q}(E)=\langle\psi_\nu|\boldsymbol\sigma\cdot\hat{\mathbf q}|\psi_\nu\rangle$ is the spin expectation value along the fixed analysis axis $\hat{\mathbf q}$. We define
\begin{equation}
P_{{\rm mode},q}(E)=\frac{1}{N_+(E)}\sum_{\nu\in\mathcal R(E)}s_{\nu q}(E),
\qquad G_{\rm ideal}(E)=\frac{e^2}{h}N_+(E).
\label{eq:ideal_mode_benchmark}
\end{equation}
Here $P_{{\rm mode},q}$ is the dimensionless right-moving-mode spin polarization, and $G_{\rm ideal}$ is the corresponding Landauer conductance for perfectly transmitting channels~\cite{Buttiker1986}. Because SOC removes a conserved up/down channel index, Eq.~\eqref{eq:ideal_mode_benchmark} is not the collinear ratio $(G_\uparrow-G_\downarrow)/(G_\uparrow+G_\downarrow)$. It is the unit-transmission limit of a Landauer spin-density-matrix readout along $\hat{\mathbf q}$~\cite{NikolicSouma2005}. The plotted curves use 5~meV Gaussian broadening of the numerator and denominator separately. Contact-dependent drain polarization and numerical details are discussed in the SM~\cite{SupplementalMaterial}.

{The spin polarization survives this channel sum in the easy-axis $[100]$ domain [Fig.~\ref{fig:transport_readout}(b)]. It reaches $P_{{\rm mode},q}\simeq-0.352$ near $E-E_{\rm ref}=16~\mathrm{meV}$ with $G_{\rm ideal}\simeq4.47\,e^2/h$, demonstrating a net polarization of about $35\%$ at finite ideal conductance. The transverse-diagonal $[1\bar10]$ domain reaches about $12.8\%$, whereas the $[001]$ and longitudinal $[110]$ domains show strongly suppressed polarization along the same detector axis. In the mirror-symmetric limit, $m_v$ forces this longitudinal mode polarization to vanish for the latter two orientations. The small residuals in the Wannier curves are discussed in the SM. Direct DFT calculations of the non-Janus Ta$_2$Te$_2$O ribbon likewise yield spin-polarized channels for the $[100]$ SOC state, with polarization $-42.9\%$ along parent $[100]$ near $E-E_F=-0.113$~eV and $G_{\rm ideal}=4e^2/h$. This benchmark uses normalized PAW spin projections rather than the full-state spin expectations in Fig.~\ref{fig:transport_readout}, with the corresponding curves provided in the SM~\cite{SupplementalMaterial}.}

{In summary, altermagnetic confinement provides a route to one-dimensional type-IV magnetism. Symmetric isolated ribbons retain nonrelativistic spin degeneracy, while SOC produces N\'eel-orientation-dependent splitting and spin-polarized propagating channels. Ta$_2$TeSeO Wannier calculations and direct Ta$_2$Te$_2$O DFT calculations establish these responses, {with the latter resolving spin-split $[100]$ and mirror-protected, spin-degenerate $[001]$ SOC spectra.} Transverse assembly of the isolated Ta$_2$Te$_2$O ribbons restores spin splitting without SOC, providing the one-dimensional counterpart of the connection between 2D type-IV layers and 3D altermagnets formed by suitable periodic stacking. The isolated-ribbon and periodic-array comparison thus links type-IV magnetism and altermagnetism through the dimensionality of electronic motion.}

\begin{acknowledgments}
We would like to thank Shandong Institute of Advanced Technology and National Supercomputing Center (Shuguang) for providing computational resources. This work is supported by the National Natural Science Foundation of Shandong Province (Grant No.~ZR2024QA040) and the National Natural Science Foundation
of China (Grant No. 12604277). Xin Chen thanks the China Scholarship Council for financial support (No.~201606220031). Duo Wang acknowledges financial support from the Science and Technology Development Fund of Macao SAR (Nos.~0062/2023/ITP2 and 0016/2025/RIA1) and Macao Polytechnic University (Grant No.~RP/FCA-03/2023).
\end{acknowledgments}

\bibliography{Ref}

\end{document}